# STDPG: A Spatio-Temporal Deterministic Policy Gradient Agent for Dynamic Routing in SDN

Juan Chen [1], Zhiwen Xiao [2], Huanlai Xing [1*], Penglin Dai [1], Shouxi Luo [1], Muhammad Azhar Iqbal [1]
[1] School of Information Science and Technology, Southwest Jiaotong University, Chengdu 611756, China
[2] Chengdu University of Information Technology, Chengdu 610225, China
* E-mail: hxx@home.swjtu.edu.cn

*Abstract*—Dynamic routing in software-defined networking (SDN) can be viewed as a centralized decision-making problem. Most of the existing deep reinforcement learning (DRL) agents can address it, thanks to the deep neural network (DNN) incorporated. However, fully-connected feed-forward neural network (FFNN) is usually adopted, where spatial correlation and temporal variation of traffic flows are ignored. This drawback usually leads to significantly high computational complexity due to large number of training parameters. To overcome this problem, we propose a novel model-free framework for dynamic routing in SDN, which is referred to as spatio-temporal deterministic policy gradient (STDPG) agent. Both the actor and critic networks are based on identical DNN structure, where a combination of convolutional neural network (CNN) and long short-term memory network (LSTM) with temporal attention mechanism, CNN-LSTM-TAM, is devised. By efficiently exploiting spatial and temporal features, CNN-LSTM-TAM helps the STDPG agent learn better from the experience transitions. Furthermore, we employ the prioritized experience replay (PER) method to accelerate the convergence of model training. The experimental results show that STDPG can automatically adapt for current network environment and achieve robust convergence. Compared with a number state-of-the-art DRL agents, STDPG achieves better routing solutions in terms of the average end-to-end delay.

*Index Terms—convolutional neural networks, deep reinforcement learning, dynamic routing, long short-term memory, software-defined networking*

## I. INTRODUCTION

By decoupling the control and data planes, software-defined networking (SDN) provides a global view for efficient network operation and management, which is beneficial to policy-based traffic delivery, fast routing response and reducing capital and operational expenditures [1]. Due to the dynamic feature of network environment, data paths for delivering traffic may vary dramatically. It is hence important to support dynamic routing when considering the practical deployment of SDN, which helps improve network resource utilization and achieve load balancing. However, to our knowledge, a number of challenges, such as traffic variation, real-time response, and intelligent decision making, need to be addressed to support fast and efficient data delivery. Dynamic routing based on machine learning has attracted increasingly more research attention [2].

Recently, deep reinforcement learning (DRL) achieves beyond human-level performance when tackling large-scale decision-making problems [3]. DRL parameterizes policies based on deep neural network (DNN), where learning agents mine complicated system states from input data (which is usually high-dimensional) and adaptively adjust policies via the experiences obtained by repeatedly interacting with target systems. Applying DRL techniques in the networking and communications domain has become an emerging trend [4][5][6]. Nevertheless, the deep Q-network (DQN) based on DRL [4] can only solve those problems with high-dimensional state space, and discrete and low-dimensional action space. In other words, it is quite difficult to find an optimum from a continuous action space. Thus, DQN cannot well adapt for a time-varying network, especially when considering dynamic routing problems in large-scale network environment.

Different from DQN, deep deterministic policy gradient (DDPG) [5] is able to learn policies from high-dimensional, continuous action space, and is thus appropriate for addressing dynamic and continuous control problems. Stampa et al. (2017) [6] and Yu et al. (2018) [7] apply DDPG agents to dynamic routing in SDN, where fully-connected feed-forward neural network (FFNN) is adopted in the actor and critic networks for feature extraction. After training, a near-optimal routing solution is generated very quickly once new routing request arrives. The two DDPG agents above offer a possibility for real-time decision making in routing.

In real-world networks, traffic matrices consist of complex spatial and temporal data [8]. It is natural to make use of them to facilitate effective routing. Unfortunately, when extracting global discriminative action policy features, FFNN does not take spatial correlation nor temporal variation of traffic flows into account. In addition, calculating and tuning huge amount of independent weights leads to significantly high computational burden, due to the full connectivity nature of FFNN.

The convolutional neural network (CNN) [9] is widely used in the area of visual processing, because of its excellent performance in spatial feature extraction. Compared with FFNN, CNN is partially-connected neural network with tied weights, which results into significantly fewer parameters for training. Actually, for an SDN network, a traffic matrix with all source-destination pairs at a certain time can be treated as an image. Hence, CNN has the potential to replace FFNN in DDPG.

CNN still belongs to feed-forward neural network (FNN), which cannot extract features related to temporal variation. The long short-term memory network (LSTM) [10], on the other hand, is a state network with memory units in favor of temporal information processing. Moreover, attention mechanisms are commonly adopted for sequence-to-sequence data analysis [11]. With these mechanisms integrated, LSTM is able to discover useful long-term dependencies.

Furthermore, in the replay buffer of the DDPG agent, uniform sampling is commonly used. However, this method probably ignores some valuable experiences in the process of sample transition. Prioritized experience replay (PER) based on DQN achieves better performance for game-playing tasks, where better experiences are sampled with higher probabilities [12].

This paper proposes a novel DRL agent, named spatio-temporal deterministic policy gradient (STDPG). Both the actor and critic networks are based on a combination of CNN and LSTM. CNN is employed to learn spatial traffic patterns from input traffic matrices. The latent features obtained are then fed into LSTM to learn valuable long-term dependencies from the traffic sequence. Through repeated interactions with dynamic network environment and frequent replay of important experiences, the STDPG agent in SDN aims at finding high-quality routing solution, where the average end-to-end delay is the objective for minimization. The experimental results show that the STDPG agent achieves not only better convergence but also more promising optimization performance than five widely used agents. Our main contributions are summarized below.

- We propose the STDPG agent for dynamic routing optimization in SDN, where both the actor and critic networks are based on CNN and LSTM for efficiently exploiting spatial correlation and temporal variation information.
- A novel temporal attention mechanism (TAM) is introduced into LSTM to emphasize on certain time steps that are key to dynamic routing decision making.
- Instead of uniform sampling, the proposed STDPG agent adopts PER in the replay buffer to make better use of high-quality experiences.

## II. RELATED WORK

Machine learning based routing has become one of the main stream research trends in SDN [2].

Yang *et al.*[13] present an approximate dynamic programming-based joint admission control and routing scheme for video streaming service. A CRS-MP routing scheme is proposed in [14], where SDN controller uses artificial neural network to learn and predict vehicle arrival rate, and makes routing decision for vehicles. However, the models above suffer from extremely high computational overhead during acquisition of labeled datasets for training.

Applying DRL techniques to routing decision making has received increasingly more research efforts. Qiu *et al.* [4] present a novel dueling DQN approach to a joint optimization problem in SDN-based IoTs, with view change, access selection and computational resource allocation taken into account. A DQN model is devised to optimize link between two unmanned air vehicles [15]. Nevertheless, DQN is not immediately applicable to continuous high dimensional action space since DQN depends on finding the best action that maximizes a given Q-value function.

On the other hand, DDPG has become a potential solution to routing optimization in SDN. DDPG-based traffic control architecture is designed for supporting multimedia communications [5]. Existing work in [6] and [7] employs DDPG to optimize routing paths for all source-destination pairs, with average end-to-end delay minimized. Experience replay techniques with uniform sampling, however, cannot achieve effective training for the DDPG agent. PER is hence introduced to a DDPG-based traffic engineering (TE) framework, where the end-to-end delay, throughput and network utility are considered [16].

Compared with DQN, DDPG agent is able to make direct use of raw observations for learning, requiring much fewer steps that learn from experiences. Nevertheless, FFNN in the DDPG agent usually involves too many parameters in the fully connected layers during training, which leads to heavy computational burden. Besides, FFNN only obtains global discriminative traffic features, ignoring spatial correlation and temporal variation. Therefore, it is in urgent need of an improved DDPG agent for highly dynamic routing problem, which motivates us to develop the STDPG agent and investigate its adaptation for the dynamic routing in SDN.

## III. SDN ROUTING FRAMEWORK BASED ON STDPG

In this paper, the routing optimization problem in SDN is regarded as a dynamic decision-making task. This section introduces the dynamic routing framework based on STDPG, as shown in Fig.1. The general process of the STDPG agent can be summarized as follows:

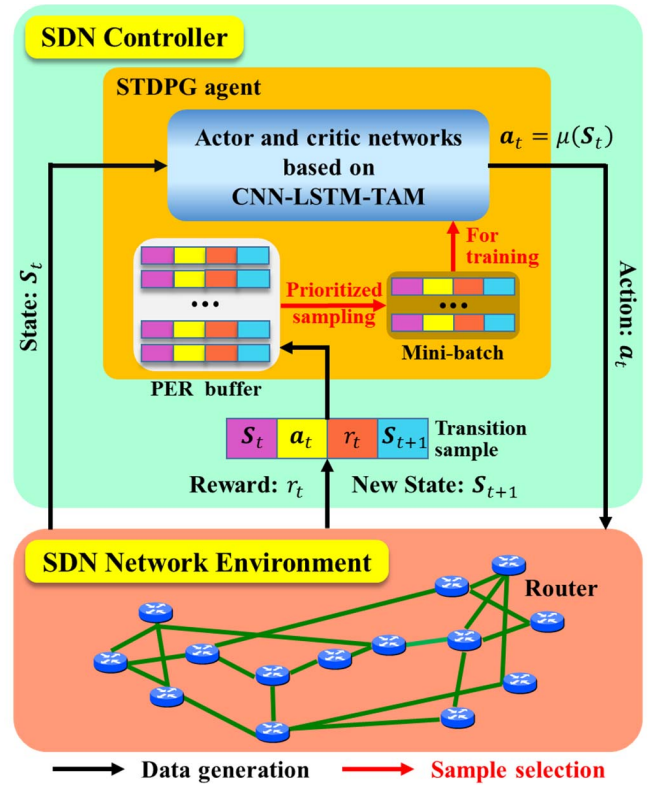

Fig. 1.   The SDN routing framework based on STDPG

First of all, the STDPG agent generates the training data. At time step *t*, this agent obtains the current network *state*, $S_t$, from the SDN controller that is periodically collected from the network environment. Let *N* denote the number of nodes in the given network. $S_t$ is the traffic matrix of the current network defined in Eq.(1), where $b_{ij}^t$ stands for bandwidth demand from node *i* to node *j*, $i,j \in \{1, …, N\}$. The STDPG agent obtains the *action* set, $a_t = \mu(S_t)$, based on state $S_t$ with deterministic policy $\mu$. $a_t$ is a solution to the dynamic routing problem in SDN, where data-paths for all source-destination pairs are included. The reward function in this paper is to minimize the average end-to-end delay. Then, the STDPG agent calculates a *reward* value based on $S_t$ and $a_t$, namely $r_t = r(S_t, a_t)$, according to Eq.(2), where $D_{ij}$ is the end-to-end delay along the path originating from node $i \in \{1, …, N\}$, and terminating at node $j \in \{1, …, N\}$. If $i = j$, we have $D_{ij} = 0$. Besides, the STDPG agent observes a new state $S_{t+1}$ by interacting with the SDN network environment. After that, the 4-tuple transition sample $(S_t, a_t, r_t, S_{t+1})$ is stored in the PER buffer.

$$S_t = \begin{pmatrix} b_{11}^t & \cdots & b_{1N}^t \\ \vdots & \ddots & \vdots \\ b_{N1}^t & \cdots & b_{NN}^t \end{pmatrix} \quad (1)$$

$$r_t = \frac{1}{N^2} \sum_{i=1}^{N} \sum_{j=1}^{N} D_{ij} \quad (2)$$

Secondly, the actor and critic networks in the STDPG agent are trained. Both of them are based on the same network structure that combines CNN and LSTM with TAM, namely CNN-LSTM-TAM. With respect to the method for training, the actor network adopts the deterministic policy gradient (DPG) algorithm while the critic network uses DQN algorithm. At each training epoch, the weights associated with the actor and critic networks are updated by sampling a mini-batch from the PER buffer, where the mini-batch is composed of a number of transition samples. The purpose for training is to get the optimal action $a_t$ based on $S_t$ and the *long-term reward* value $R = \sum_{t=0}^{T} \gamma^t \cdot r(S_t, a_t)$. The discount factor $\gamma$ ($0 \leq \gamma < 1$) is a constant indicating the importance of the reward value obtained in the future. *T* is the predefined number of time steps. The training process is iteration-based, where quality of the routing solution is gradually improved.

Last but not least, the STDPG agent is ready for use after training. When a new routing request (traffic matrix) arrives, near-optimal data forwarding paths are rapidly generated in a single step.

## IV. DEEP LEARNING MODEL DESIGN FOR STDPG

The traditional DDPG agent with FFNN is not stabilized due to that the agent only exploits the current state and extracts global discriminative features, with spatial correlation and temporal variation ignored. However, traffic matrices in SDN network are complex spatial-temporal data that is not suitable for FFNN-based models.

In this section, we introduce a CNN-LSTM-TAM model to replacing the FFNN-based model in both the actor and critic networks, as shown in Fig.2.

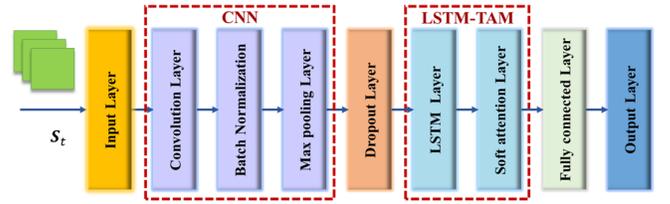

Fig. 2. The CNN-LSTM-TAM model

### A. The CNN-LSTM-TAM Model

In the STDPG agent, both the actor and critic networks employ identical CNN-LSTM-TAM model. Yet, they have different input and output. To be specific, at each time step *t*, the actor network takes traffic matrix $S_t$ and routing solution $a_t$ as its input and output data, respectively. On the other hand, based on $S_t$ and $a_t$ as its input data, the critic network outputs a value function $Q(S_t, a_t)$ that is the expected discounted cumulative reward defined in Eq.(3).

$$Q(S_t, a_t) = \mathbb{E}[R_t] \quad (3)$$

where, $R_t = \sum_{i=t}^{T} \gamma^i \cdot r(S_t, a_t)$.

In Fig. 2, the input traffic matrix is sampled by window-sliding along the temporal direction. CNN is employed to learn the spatial traffic patterns of the input. Its structure depends on the size of the convolution kernel and the sub-sampling factor. In CNN, the max pooling layer is responsible for decreasing the dimension size of features. The batch normalization layer is applied to prevent gradient explosion. The latent features captured by CNN at time step *t* are then fed into LSTM with TAM for meaningful sequence learning.

For the LSTM layer, its input and output are recurrent over time steps. At time step *t*, the input is the latent features from CNN, the last hidden state $h_{t-1}$ and the cell state $c_{t-1}$, while the output is the hidden state $h_t$ and the cell state $c_t$. Both $h_t$ and $c_t$ are updated and then input to the LSTM network at each time step. The soft attention service is to find more valuable long short-term dependencies. Then, the embedding features obtained by the soft attention layer are input to the fully connected layer for manifest expression. An action is generated as the output at time step *t*.

### B. Temporal Attention Mechanism (TAM)

LSTM implicitly utilizes the subsequent traffic sequence information contained in its hidden states to extract features. However, the traditional LSTM cannot make full use of the hidden states. How to effectively capture subsequent traffic sequence information is still a challenge.

In this paper, we employ attention mechanism to focus on the salient parts of a traffic sequence, where the extracted features are further classified along the temporal direction [11]. For time step *t,* we define an input traffic sequence $H_t \in \mathbb{R}^{N \times d_{model}}$, where *N* is the total number of source-destination pairs, and $d_{model}$ is the dimension of the embeddings within TAM. Given input query $U$, key $E$ and value $V$ matrices, the matrix of output is computed by Eq. (4).

$$\text{Attention}(U, E, V) = softmax\left(\frac{UE^T}{\sqrt{d_e}}\right) \cdot V \quad (4)$$

where, $U = H_t W^U$, $E = H_t W^E$, and $V = H_t W^V$. $W^U \in \mathbb{R}^{d_{model} \times d_e}$, $W^E \in \mathbb{R}^{d_{model} \times d_e}$, and $W^V \in \mathbb{R}^{d_{model} \times d_v}$ are weight matrices, respectively. Besides, $d_e$ is the dimension of $U$ and $E$, and $d_v$ is the dimension of $V$.

*C. Prioritized Experience Replay (PER)*

In the traditional DDPG, uniform sampling is applied to the experience replay buffer to obtain a set of transition samples. This approach treats all experiences equally, with their significance ignored. However, more important experiences usually have greater impact on the actor and critic networks with respect to the convergence, which helps reduce the number of training epochs and thus the computational burden incurred. Fortunately, the PER method are priority-based, where more important experiences are more likely to be sampled [12][16].

Instead of uniform sampling, this paper applies PER to the buffer of the STDPG agent. Let $D_t$ and $\nabla Q(S_t, a_t)$ denote the temporal-difference (TD) error and the $Q$ gradient of the actor and critic networks at time step *t*, respectively.

The actor and critic networks are trained together based on the transition samples obtained by the PER buffer. The priority value of the transition sample at time step *t*, $v_t$, is defined in Eq. (5), where $\alpha$ ($0 < \alpha < 1$) is a constant implying the relative importance of TD error against the $Q$ gradient, and $\varepsilon$ is a small positive constant for avoiding edge-cases that a transition is not revisited once TD error is zero.

$$v_t = \alpha(|D_t| + \varepsilon) + (1 - \alpha)|\nabla Q(S_t, a_t)| \quad (5)$$

Let $v_t^k$ be the *k*-th transition sample in PER buffer at *t*. The replay probability of the *k*-th transition in the buffer, $p_t^k$, is defined in Eq. (6)▷

$$p_t^k = \frac{(v_t^k)^\beta}{\sum_{k=1}^{M}(v_t^k)^\beta} \quad (6)$$

where, $\beta$ ($0 \leq \beta \leq 1$) is a constant controlling the contribution of the prioritization and *M* is the buffer size.

## V. PERFORMANCE EVALUATION

In this section, we first introduce the experimental setup and then discuss the results and analysis in detail.

*A. Experimental Setup*

The experimental network topology is GEANT2 with 24 nodes and 37 full-duplex links. Intensity level of traffic (ILT) is defined as a percentage indicating how much the total network bandwidth is consumed. Different ILT values can mimic real-world network environment in terms of the traffic. We set five ILT values in the experiments, namely, 20%, 40%, 60%, 80%, and 100%. Note that for each ILT, all traffic sequences are generated based on the gravity model [17].

We set up the SDN network environment by OMNET++. The proposed STDPG agent interacts with OMNET++ to simulate packet-in and packet-out scenarios. The CNN-LSTM-TAM models for the actor and critic networks are implemented by TensorFlow. Adam Optimizer is used to optimize the loss function during training. We run and train the STDPG agent on a workstation with Intel Xeon E5-2667V4 8Core CPU × 2 @3.2GHz, 128 GB RAM, and 4×NVIDIA GTX Titan V 12G GPU.

As mentioned in Subsection IV.A, both the actor and critic networks are based on identical CNN-LSTM-TAM model. In this model, the input size of the state (i.e. traffic matrix of all source-destination pairs) is 23×24. The matrix is not symmetric because we assume any source does not send data to itself. The CNN convolutional layer is composed of 64 filters. For each filter, the kernel size is 2×2 and the stride size is 1. For the max-pooling layer, the pooling and stride sizes are set to 2×2 and 1, respectively. In terms of regularization, we employ dropout operation with a dropout rate of 0.5. The LSTM layer and the output layer have 64 hidden units, respectively, where the latter is a fully connected layer.

The structural parameters of the CNN-LSTM-TAM model are shown in Table I. The out shape stands for dimension transformation associated with a given layer. The number of parameters represents that of weight vectors in each layer. There are only 84234 trainable parameters in total. Hence, it is easy to run and train the proposed STDPG agent.

Other important parameters of the STDPG agent are set below. The learning rates of the actor and critic networks are initially set as 0.001 and 0.0001. We set the discount factor of the critic network and the update rate of target networks to 0.99 and 0.001, respectively. The size of mini-batch is set as 32. The PER buffer is designed as a circular queue with a length of 2000.

*B. Results and Analysis*

In this subsection, we first evaluate the performance of the proposed STDPG agent with respect to the convergence, where the average end-to-end delay is the objective (reward function) for minimization. We then compare performance of six DRL-based agents.

The first experiment is to verify the convergence of the STDPG agent during training. For each ILT, we generate 1000 different traffic sequences and set the predefined numbers of training epochs to 10, 20, 30, 40, 50 and 60, respectively. There are 1000 time steps in each training epoch. We collect the average end-to-end delay values of each ILT instance in Fig.3.

Firstly, one can easily observe that the average end-to-end delay steadily grows up, with ILT increasing. A larger ILT leads to a higher average end-to-end delay. The following explains why. With the total bandwidth resource limited, the more traffic flows are poured into the network, the less bandwidth is allocated to each of them and the longer transmission time is hence needed for delivering each flow. Simultaneously handling more traffic flows at each intermediate node leads to higher processing delay and thus much higher average end-to-end delay along each data-path.

TABLE I  HYPER-PARAMETERS OF CNN-LSTM-TAM MODEL IN ACTOR AND CRITIC NETWORKS

| Construction | Actor network | | Critic network | | | |
| --- | --- | --- | --- | --- | --- | --- |
| | Out shape | No. of Parameters | Out shape | | No. of Parameters | |
| | state | state | state | action | state | action |
| Input Layer | (None,23,24) | 0 | (None,23,24) | (None,1,37) | 0 | 0 |
| Convolution Layer | (None,22,23,64) | 320 | (None,22,23,64) | (None,1,36,64) | 320 | 192 |
| Batch Normalization | (None,22,23,64) | 256 | (None,22,23,64) | (None,1,36,64) | 256 | 256 |
| Max pooling Layer | (None,11,11,64) | 0 | (None,11,11,64) | (None,1,18,64) | 0 | 0 |
| Dropout Layer | (None,121,64) | 0 | (None,121,64) | (None,18,64) | 0 | 0 |
| LSTM Layer | (None,64) | 33024 | (None,64) | (None,64) | 33024 | 33024 |
| Soft attention Layer | (None,64) | 0 | (None,64) | (None,64) | 0 | 0 |
| Fully connected Layer | (None,64) | 4160 | (None,64) | | 5888 | |
| Output Layer | (None,37) | 2405 | (None,37) | | 2405 | |

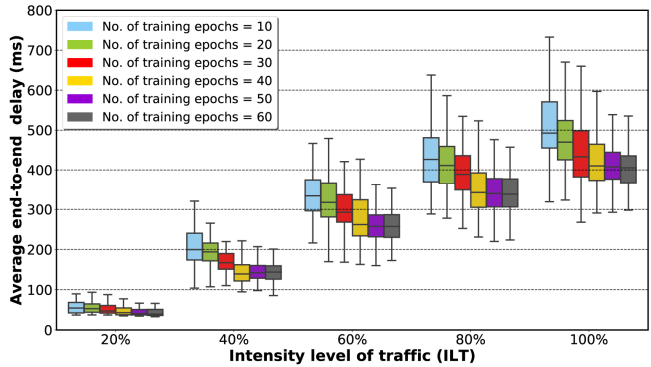

Fig. 3.  STDPG agent learning process with different ILTs

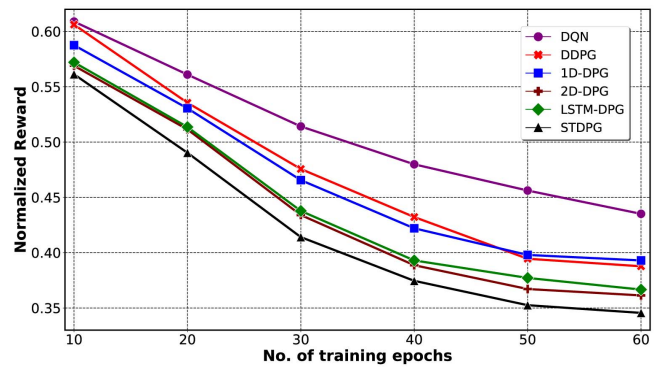

Fig. 4.  Reward values obtained during training

Secondly, if emphasizing on each ILT instance, one can clearly see that with the number of training steps increasing, the average end-to-end delay value gradually goes down. In particular, the upper and lower quartile values of each boxplot tend to draw close to each other, from the 40th epoch to the 60th epoch, indicating the CNN-LSTM-TAM model can converge within 60 epochs.

The second experiment evaluates the effectiveness of the STDPG agent. We compare it with five state-of-the-art DRL agents, including DQN [15], DDPG [5], DPG based on 1D CNN [9] (1D-DPG), DPG based on 2D CNN [18] (2D-DPG) and DPG based on LSTM (LSTM-DPG) [19]. For fair comparison, all DPG agents are exactly the same except for the DNN model used in actor and critic networks. Besides, the DQN and DDPG agents are based on the same FFNN model.

Firstly, we compare the reward values obtained by the six agents during training. For simplicity purpose, we use the normalized reward $r_{normalized}(S,a)$ defined in Eq. (7), where $r(S,a)$, $r_{min}(S,a)$ and $r_{max}(S,a)$ stand for the average, minimum and maximum rewards, respectively.

$$r_{normalized}(S,a) = \frac{r(S,a) - r_{min}(S,a)}{r_{max}(S,a) - r_{min}(S,a)} \quad (7)$$

Due to page limit, we take ILT = 80% as an example and show the normalized reward values in Fig.4. Clearly, STDPG delivers the best performance among the six agents, which demonstrates the effectiveness of the STDPG agent with the CNN-LSTM-TAM model and PER. Actually, we observe quite similar phenomenon with different ILTs.

Secondly, we compare the average end-to-end delay values obtained by the six agents after training. In each ILT instance, we generate 100 different traffic sequences. Note that violin plot is a combination of box plot and kernel density plot [20]. Fig.5 shows the violin plots of the six agents regarding the average end-to-end delay, where the outer shape of each violin implies the density estimation of the corresponding average end-to-end delay values.

We have two findings by observing Fig.5. First of all, for each agent, heavier traffic in the network leads to higher average end-to-end delay values, as explained before. In each instance, the STDPG agent achieves the best median value among all agents. Secondly, for each agent, with the amount of traffic in the network growing up, the distribution of the average end-to-end delay values becomes more and more dispersed. Similarly, the STDPG agent performs the best.

In addition, we collect the median values of the average end-to-end delays (*ms*) produced by each DRL agent in Table II. Obviously, STDPG outperforms DQN, DDPG, 1D-DPG, 2D-DPG as well as LSTM-DPG as it always obtains the smallest median value in each instance. For example, when ILT = 100%, STDPG outperforms DQN, DDPG, 1D-DPG, 2D-DPG and LSTM-DPG with respect to the median value by 30.92%, 22.51%, 20.32%, 10.88% and 12.81%, respectively.

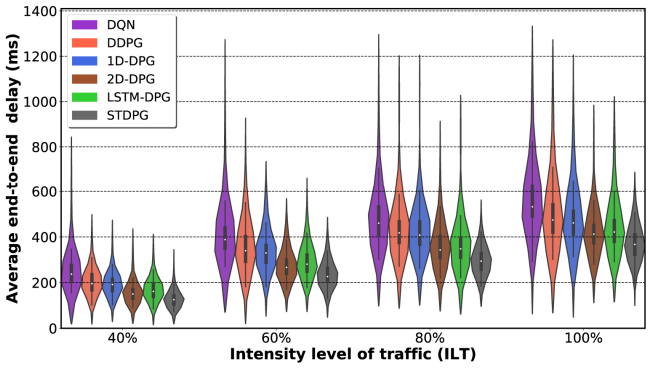

Fig. 5. Comparisons of the average end-to-end delay

TABLE II  RESULTS OF MEDIAN VALUES (MS)

| Agents | ILT=20% | ILT=40% | ILT=60% | ILT=80% | ILT=100% |
|---|---|---|---|---|---|
| DQN | 70.15 | 237.30 | 388.50 | 460.20 | 532.05 |
| DDPG | 53.10 | 195.53 | 339.96 | 417.99 | 474.32 |
| 1D-DPG | 52.55 | 191.65 | 329.95 | 408.05 | 461.30 |
| 2D-DPG | 45.70 | 150.50 | 266.70 | 343.00 | 412.40 |
| LSTM-DPG | 48.10 | 161.59 | 280.75 | 347.93 | 421.57 |
| STDPG | **41.35** | **125.25** | **226.55** | **294.05** | **367.55** |

According to the above observations, it is no doubt that the STDPG agent achieves the best overall performance in terms of the average end-to-end delay. The following explains why. On the one hand, the action space of DQN is discrete and thus inappropriate for continuous control. DDPG is able to handle continuous control problems since it maps state space to action space via DNN. However, the FFNN model used in the DDPG agent ignores the spatio-temporal information of traffic data as it only extracts global discrete features. Based on 1D CNN, the 1D-DPG agent makes use of temporal features. Unfortunately, it is unable to mine spatial correlation nor capture long-term change of traffic sequence. Although 2D CNN can extract spatial features, it is still FNN in nature, like FFNN and 1D CNN, which cannot learn long-range representations from traffic data effectively. LSTM-DPG achieves decent routing solutions compared with DQN, DDPG, and 1D-DPG. Nevertheless, it is more efficient for time-series problems as the LSTM model makes better use of temporal variation than spatial correlation. On the other hand, the CNN-LSTM-TAM model is able to not only learn spatial and temporal traffic patterns, but also capture long-term dependencies. By keeping learning runtime dynamics and making wise decisions towards optimal routing, the STDPG agent delivers the best overall performance with the help of CNN-LSTM-TAM and PER.

## VI. CONCLUSION

In this paper, we present a model-free end-to-end learning framework to dynamic routing in SDN, i.e. spatio-temporal deterministic policy gradient (STDPG) agent, where the objective is to minimize the average end-to-end delay. Both the actor and critic networks are based on a CNN-LSTM-TAM model that is able to capture spatial and temporal traffic patterns. Meanwhile, we apply PER to differentiate the contributions from different experiences, where experiences with higher significance are selected with higher probabilities. The experimental results show that the proposed STDPG agent outperforms five DRL agents based on FFNN, CNN or LSTM, in terms of the average end-to-end delay.